# Interpretable machine learning for identifying individual-specific cardiogram signatures

Ilija Tanasković[a,b]*
Ljiljana B. Lazarević[c]
Goran Knežević[d]
Nikola Milosavljević[e]
Olga Dubljević[f]
Bojana Bjegojević[c,1]
Nadica Miljković[a,h]

[a]University of Belgrade – School of Electrical Engineering, Bulevar kralja Aleksandra 73, 11000 Belgrade, Serbia
[b]Institute for Artificial Intelligence Research and Development of Serbia, Fruškogorska 1, 21000 Novi Sad, Serbia
[c]Institute of Psychology and LIRA lab, Faculty of Philosophy, University of Belgrade, Čika Ljubina 18-20, 11000 Belgrade, Serbia
[d]Department of Psychology and LIRA lab, Faculty of Philosophy, University of Belgrade, Čika Ljubina 18-20, 11000 Belgrade, Serbia
[e]University of Belgrade – Faculty of Special Education and Rehabilitation, Visokog Stevana 2, 11000 Belgrade, Serbia
[f]Institute for Biological Research ”Siniša Stanković”, National Institute of the Republic of Serbia, University of Belgrade, Bulevar Despota Stefana 142, 11000 Belgrade, Serbia
[1]Present address: Trinity College Dublin - School of Psychology, Aras an Phiarsaigh, College Green, Dublin 2, D02 PN40, Ireland
[h]Faculty of Electrical Engineering, University of Ljubljana, Tržaška cesta 25, 1000 Ljubljana, Slovenia

## Abstract

This study investigates cardiogram-based biometric identification as a physiological authentication problem, focusing on the discriminative capacity of electrocardiogram (ECG) and impedance cardiogram (ICG) features and their robustness to emotional variability. A total of 29 features spanning four domains (temporal, amplitude, slope, and morphological) are evaluated using random forest (RF) models combined with multiple interpretability methods. Features derived from ECG and ICG rank consistently among the top 10 most important by Gini importance, permutation importance, and Shapley additive explanations (SHAP) values, with ECG-derived QRS descriptors occupying the highest positions. In parallel, BCX features derived from ICG provide complementary information, although their stability across methods is lower. Correlation analysis reveals substantial multicollinearity, where the RF distributes and diminishes importance across highly correlated pairs, confirming reduced independent contributions. Statistical analysis identifies 14 features that differ significantly between baseline and anger, without a clear pattern by domain. Feature selection with recursive feature elimination and genetic algorithms converges on a subset (12 features) that attains accuracy within 1% of the full set (99%), improving efficiency in storage and computation. Complementary analyses indicate that individual differences are primarily encoded in ECG features describing QRS amplitude, slope, morphology, and duration, with the first three remaining stable between baseline and anger, and duration being the only selected QRS feature that changes significantly. BCX amplitude features provide supportive but less stable discriminatory cues. More broadly, the proposed framework may help distinguish stable person-related characteristics, such as identity and personality-related traits, from transient affective changes.

*Corresponding Author: Tel: +381 11 3218 348, E-mail: ilija.tanaskovic@ivi.ac.rs (Ilija Tanasković)

Other e-mail addresses: ljiljana.lazarevic@f.bg.ac.rs (Ljiljana B. Lazarević), gknezevi@f.bg.ac.rs (Goran Knežević), nikolamilosavljevic@fasper.bg.ac.rs (Nikola Milosavljević), olga.dubljevic@ibiss.bg.ac.rs (Olga Dubljević), bojanazb@tcd.ie (Bojana Bjegojević), nadica.miljkovic@etf.bg.ac.rs (Nadica Miljković)

## 1. Introduction

Physiological and behavioral characteristics can carry stable individual-specific information, allowing identity to be established or verified through measurable human characteristics (Jain et al., 2007). While fingerprints and iris patterns remain the accepted standard, recent advances highlight the value of biomedical signals, which are harder to steal or forge. Among these, the electrocardiogram (ECG) has emerged as a promising modality, due to its inherent liveness detection capability (Odinaka et al., 2012; Pereira et al., 2023; Srivastva et al., 2022). Additionally, multimodal approaches, combining ECG with other biomedical signals, have attracted increasing interest (Pereira et al., 2023). One such signal is the impedance cardiogram (ICG), which captures the changes in thoracic impedance linked to the mechanical activity of the heart (Cacioppo et al., 2007; Sherwood et al., 1990), and has shown additional value for biometric identification when integrated with ECG (Rathore et al., 2021; Tanasković et al., 2024).

The rapid adoption of machine learning (ML) has further advanced the field of biometric identification, consistently achieving high accuracy with cardiogram-based systems (Odinaka et al., 2012; Pereira et al., 2023). As predictive performance has steadily improved, however, a different challenge has become increasingly important: understanding why these systems work. Most existing research has focused on predictive performance rather than on identifying the discriminative contribution of specific ECG and ICG features. This gap restricts insight into the physiological mechanisms that underlie identity-specific signatures in healthy individuals, while also limiting the transparency of model decisions. This issue is broader than biometric identification alone, because ECG-derived features may also reflect relatively stable psychological characteristics, suggesting that cardiographic signals can contain multiple layers of person-related information (Boljanić et al., 2022). Therefore, identifying the physiological carriers of individual-specific information is relevant not only for authentication, but also for understanding how stable traits and transient states are represented in cardiographic signals. From an applied perspective, such interpretation is important because users may differ in their cognitive strategies, expectations, and interaction with authentication technologies (Belk et al., 2017). Accordingly, interpretable cardiogram-based models may support both clearer decision logic in physiological authentication and broader psychological applications in which the origin of person-related signal patterns needs to be understood.

Furthermore, cardiogram-based biometrics face several real-world challenges, including inconsistency over time and susceptibility to emotional states (Pereira et al., 2023). As the autonomic nervous system modulates cardiac function, emotional arousal can alter ECG and ICG morphology (Meltzer & Luengo, 2025; Pereira et al., 2023). This creates an important interpretive problem, because cardiographic signals may simultaneously carry stable identity- or trait-related information and transient affective or autonomic variability. Notably, emotional effects are not uniform across affective conditions, previous results obtained on the same dataset showing that anger reduces identification accuracy (Tanasković et al., 2024). A comparable decrease in cardiogram-based biometric performance was also reported for fear and happiness conditions, although this effect may partly reflect other session-dependent factors (Carvalho & Brás, 2024). In contrast, anxiety (Israel et al., 2005) and psychological stress (Zhou et al., 2021) have been reported to exert a smaller influence. Since ECG and related physiological signals can reflect emotional responses during technology-mediated tasks and complement subjective assessment of user state (Li et al., 2025; Snow et al., 2025; Wadsley & Ihssen, 2025), cardiographic recordings may contain both identity-related and affect-related information. Therefore, emotionally sensitive features must be distinguished from robust identity cues. The present study extends previous findings to the feature level, aiming to determine whether performance degradation under anger is driven by changes in the most identity-informative descriptors or by less discriminative, emotion-sensitive components. This is necessary for understanding how physiological carriers of identity behave under affective variation and for developing models whose decisions are both interpretable and practically robust.

## 1.1 Research questions

Previous studies show that (Antić et al., 2020; Tanasković et al., 2024) demonstrated that a multimodal approach utilizing fiducial-based features extracted from both ECG and ICG signals achieved the highest biometric identification accuracy, with statistically significant results, compared with the single-signal approach. However, those efforts focused mainly on predictive performance, rather than on understanding which physiological features carry individual-specific information and whether those features remain stable across emotional states.

In this study, we therefore adopt an exploratory approach (Shmueli, 2010; Yarkoni & Westfall, 2017), aimed at identifying the physiological characteristics that encode individual-specific information in cardiographic signals, including heartbeat amplitude, waveform morphology, and slope, or the variability of heart intervals. Accordingly, the study is guided by the following research questions:

1. Which components of the cardiogram signals contain the most informative physiological features for biometric identification?
2. How does correlation among features influence their importance in ML models, such as random forests (RF)?
3. Which cardiographic features are the most affected by emotional state changes, and do particular feature types (*e.g.,* time, amplitude, slope, or morphology) exhibit consistent sensitivity to emotional variations?

# 2. Materials and methods

The publicly available dataset of simultaneously recorded ECG and ICG signals from 202 healthy young adults (Tanasković et al., 2023, 2024) is used for the purpose of this study, with full details on the dataset, recording setup, and acquisition procedures provided in the Supplementary materials (Tanasković et al., 2025). The visual representation of the proposed methodology is shown in Figure 1. The pipeline begins with preprocessing to reduce noise, followed by delineation to identify fiducial points, *i.e.*, characteristic landmarks tied to key cardiovascular events. These two steps follow previously described procedures (Antić et al., 2020; Tanasković et al., 2024). An RF is initially trained and evaluated using all 29 features, which serves as the base model and benchmark, with identification accuracy used as the reference metric for all subsequent comparisons.

Several techniques are then applied in parallel to identify a specific subset of features that are crucial for identification, following an ensemble feature selection approach (Bolón-Canedo & Alonso-Betanzos, 2019). Feature importance is estimated with three methods (Gini importance, permutation feature importance, and shapley additive explanations - SHAP values), from which the top 10 features are taken and intersected to produce the first subset, providing a stable consensus ranking. Second, feature selection is performed using recursive feature elimination with cross-validation (RFECV) and a genetic algorithm (GA), each producing a data-driven subset without manual thresholds. The intersection of these two subsets defines the second subset, which emphasizes selection stability. Third, correlation analysis is conducted using Pearson and Spearman coefficients to identify clusters of highly correlated features, which are then used to assess cluster-level contribution and to examine how multicollinearity influences importance estimation. Finally, statistical testing across emotional states identifies features that change significantly, and this inference is complemented by training the RF on three subsets: all features, the non-significant set, and the significant set, to compare predictive behavior. All evaluations are conducted on an independent test set, which has not participated in the training process. The outcomes from these four paths are synthesized to localize where identity-related information resides in ECG and ICG and to characterize how emotional variability shapes those signals. To support clarity and ease of navigation, the detailed steps of the proposed pipeline are provided in the Supplementary materials (Tanasković et al., 2025).

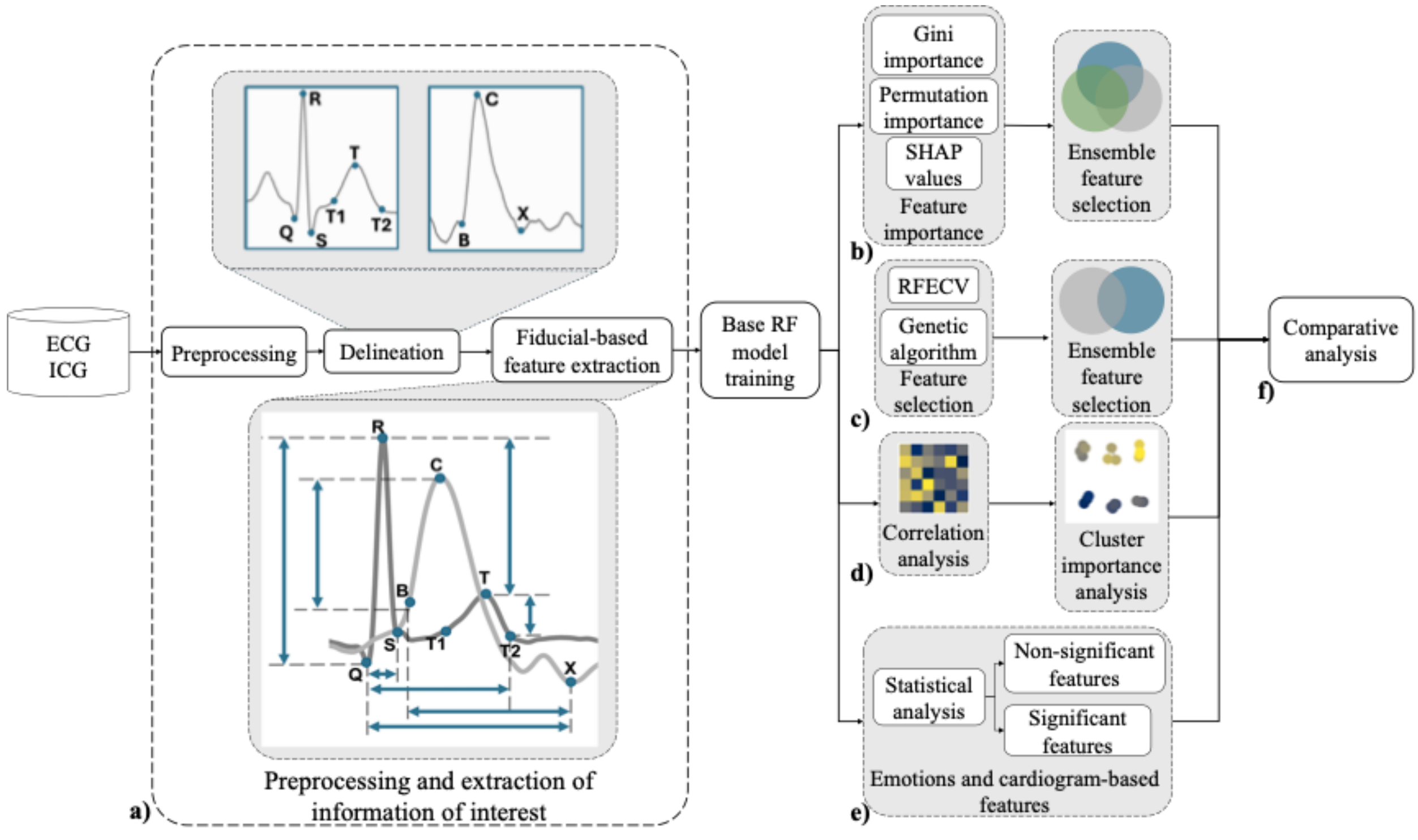

*Figure 1: Block diagram of the proposed methodology: a) preprocessing and extraction of information of interest; b) feature importance; c) feature selection; d) correlation and cluster-importance analysis; e) emotions and cardiogram-based features; f) comparative analysis. electrocardiogram - ECG, impedance cardiogram - ICG, random forest - RF, shapley additive explanations - SHAP, recursive feature elimination with cross-validation - RFECV.*

## 3. Results

Feature cross-correlations are visualized as heatmaps, with significant associations ($p < 0.05$) marked by an asterisk. As Pearson and Spearman matrices show nearly identical patterns, only Pearson coefficients ($r$) are presented in Figure 2, while the Spearman matrix ($\rho$) is provided in Figure A1 of the Supplementary Materials (Tanasković et al., 2025). Highly correlated pairs ($|r|$ or $|\rho| > 0.7$) are summarized in Table 1. All reported correlations are significant ($p < 0.001$) and each pair identified by Spearman is also captured by Pearson correlation coefficient. Therefore, subsequent analysis relies on Pearson correlations. Using a graph search over the adjacency list of these pairs, seven clusters of highly correlated features are identified for further analysis.

### 3.1 Feature importance findings

The RF model achieves 99.17% accuracy, 99.16% F1 score, 99.33% precision, and 99.17% recall on the test set using all features, described in Table A1 provided in the Supplementary materials (Tanasković et al., 2025). These results are referred to as the base model performance in subsequent comparisons. Detailed feature importances are provided in Figures A2-A4 in the Supplementary Materials (Tanasković et al., 2025).

Using an ensemble feature-selection strategy, the intersection of the three top 10 lists yields a consensus subset of 8 features, visualized as a Venn diagram in Figure 3, consisting of ECGQRScrest, RS_amp, RS_slope, RT_amp, TT2_amp, QRS_int, RQ_amp, and QR_slope. Retraining the RF model with this subset results in 96.92% accuracy, 96.91% F1 score, 96.46% precision, and 96.92% recall.\

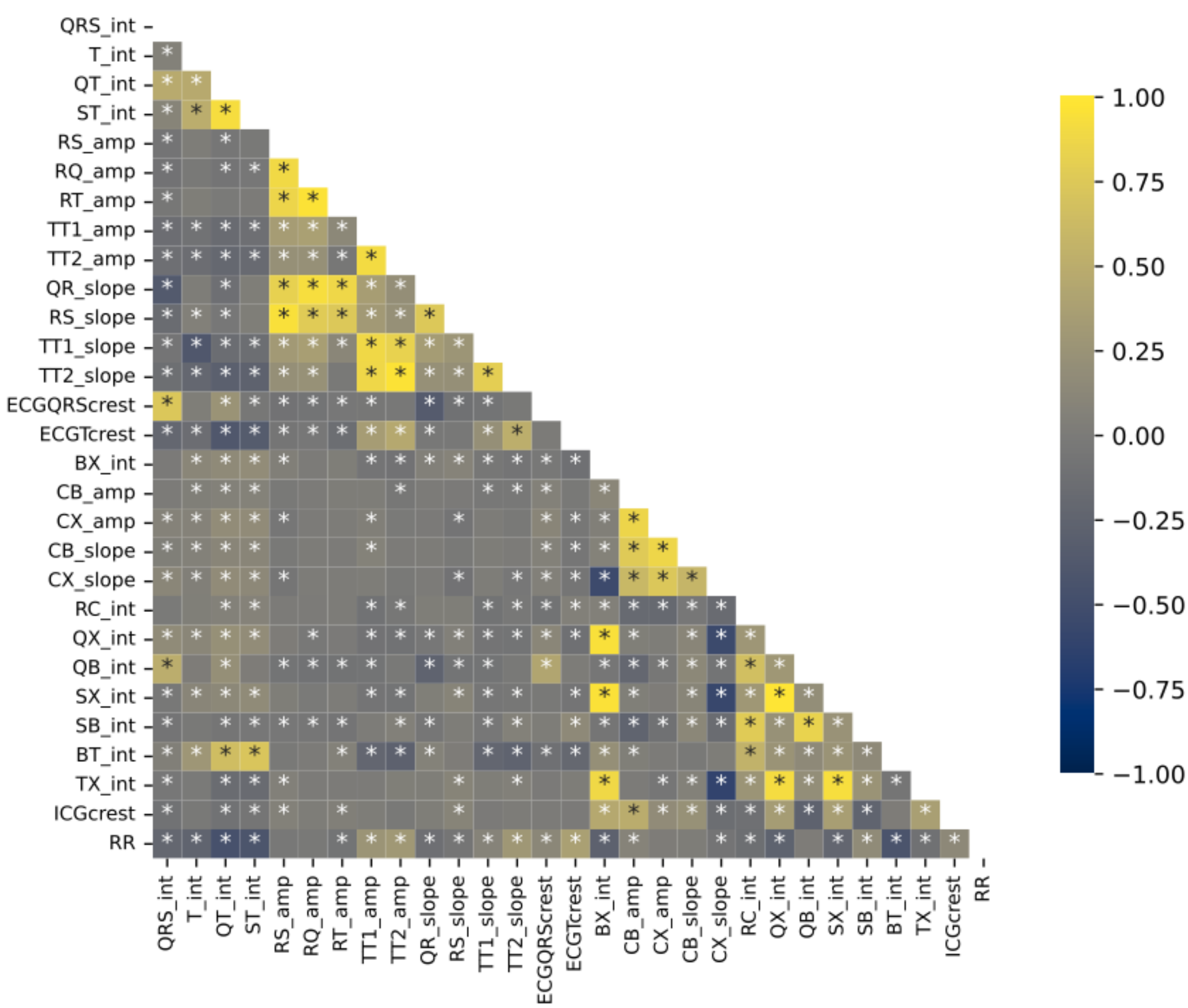

*Figure 2: Heatmap of Pearson correlation coefficients between extracted features, with asterisks (*) indicating statistical significance.*

### 3.2 Correlation and cluster analysis outcomes

The effect of multicollinearity on feature importance is summarized in Table 2, which reports the seven identified feature clusters, while their collective contribution to classification is visualized in Figure 4. For each feature (Variable 1), a permutation is applied to quantify the relative change in the importance of its correlated pair (Variable 2), while the resulting changes in importance across all other features are summarized using median and interquartile range (IQR).

### 3.3 Feature selection performance

RFECV selects 15 features, whereas the GA selects 17. Both methods achieve very high performance, with RFECV reaching 99.00% accuracy and F1 score, 99.18% precision, and 99.00% recall, and the GA achieving 98.58% accuracy, 98.56% F1 score, 99.81% precision, and 98.58% recall. Despite differences between the subsets, there is substantial overlap, with both methods consistently selecting 12 features. The visualization of the overlapping and unique features is shown in Figure 5. The RF model is retrained using the 12 intersected features, achieving 98.17% accuracy, 98.15% F1-score, 98.45% precision, and 98.15% recall, which is comparable to the base model, despite reducing the feature set by more than half.

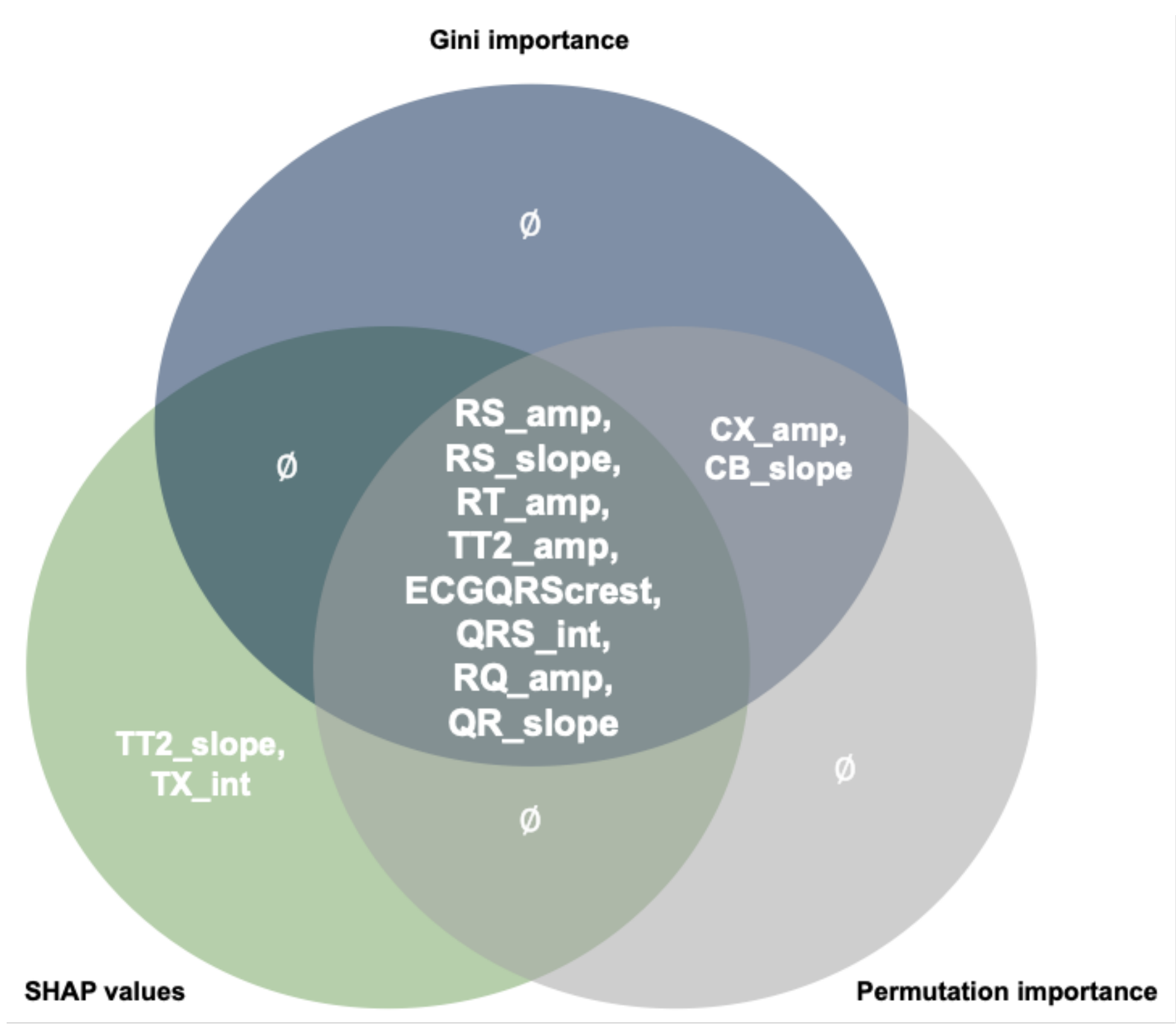

*Figure 3: Venn diagram showing the overlap among the top 10 features identified by Gini, permutation importance, and SHAP methods.*

### 3.4 Effect of emotions on cardiogram-based features

Statistical analysis reveals that 14 features (QRS_int, T_int, QT_int, ST_int, TT1_amp, TT1_slope, BX_int, CB_amp, CB_slope, RC_int, QX_int, BT_int, ICGcrest, and RR) differ significantly between the baseline and anger segments. The remaining 15 features, such as RS_amp, RQ_amp, RT_amp, TT2_amp, QR_slope, RS_slope, TT2_slope, ECGQRScrest, ECGTcrest, CX_amp, CX_slope, QB_int, SX_int, SB_int, and TX_int, do not exhibit significant changes across emotional states. Detailed results of statistical analysis are presented in Table A2 in the Supplementary Materials (Tanasković et al., 2025). The effect sizes are negligible for all features except T_int, which shows a small effect. The highly correlated pairs that show opposite significance patterns between segments are presented in Table 3.

*Table 1: Pairs of correlated features, denoted as Feature 1 and Feature 2, with the corresponding Pearson and Spearman correlation coefficients; the Method column indicates whether the high correlation (*$|r|$ *or* $|\rho| > 0.7$*) is identified by Pearson only, Spearman only, or by both. Suffixes in the feature names indicate the feature domain: _int denotes the temporal domain, comprising interval and duration measures, _amp denotes the amplitude domain, and _slope denotes the slope domain. All correlations are statistically significant with* $p < 0.001$*.*

| Feature 1 | Feature 2 | Pearson | Spearman | Method |
|---|---|---|---|---|
| QX_int | SX_int | 0.981 | 0.971 | Both |
| RQ_amp | RT_amp | 0.958 | 0.949 | Both |
| TT2_amp | TT2_slope | 0.954 | 0.964 | Both |
| BX_int | SX_int | 0.953 | 0.954 | Both |
| RS_amp | RS_slope | 0.943 | 0.942 | Both |
| BX_int | QX_int | 0.939 | 0.937 | Both |
| SX_int | TX_int | 0.926 | 0.902 | Both |
| RQ_amp | QR_slope | 0.922 | 0.918 | Both |
| QT_int | ST_int | 0.918 | 0.846 | Both |
| QX_int | TX_int | 0.907 | 0.873 | Both |
| TT1_amp | TT2_amp | 0.898 | 0.903 | Both |
| RS_amp | RQ_amp | 0.891 | 0.866 | Both |
| TT1_amp | TT1_slope | 0.881 | 0.874 | Both |
| BX_int | TX_int | 0.876 | 0.857 | Both |
| TT1_amp | TT2_slope | 0.874 | 0.888 | Both |
| RT_amp | QR_slope | 0.871 | 0.858 | Both |
| CX_amp | CB_slope | 0.865 | 0.903 | Both |
| RS_amp | RT_amp | 0.857 | 0.833 | Both |
| CB_amp | CX_amp | 0.852 | 0.835 | Both |
| QB_int | SB_int | 0.830 | 0.710 | Both |
| TT2_amp | TT1_slope | 0.821 | 0.830 | Both |
| RS_amp | QR_slope | 0.816 | 0.793 | Both |
| TT1_slope | TT2_slope | 0.792 | 0.799 | Both |
| RC_int | SB_int | 0.770 | 0.271 | Pearson only |
| RQ_amp | RS_slope | 0.770 | 0.748 | Both |

| Feature 1 | Feature 2 | Pearson | Spearman | Method |
|---|---|---|---|---|
| CB_amp | CB_slope | 0.764 | 0.817 | Both |
| QR_slope | RS_slope | 0.739 | 0.721 | Both |
| RT_amp | RS_slope | 0.735 | 0.714 | Both |
| QRS_int | ECGQRScrest | 0.734 | 0.391 | Pearson only |
| CX_amp | CX_slope | 0.731 | 0.676 | Pearson only |
| ST_int | BT_int | 0.712 | 0.649 | Pearson only |

The predictive evaluation of features, based on sensitivity to emotions, is summarized in Table 4. RF models are trained using three feature subsets: all features, only emotion-insensitive (non-significant) features, and only emotion-sensitive (significant) features. Utilizing all features, within-segment performance remains high (97–99%), while cross-segment generalization declines (88–89% baseline → anger; 91–92% anger → baseline). Using only emotion-insensitive features preserves strong within-segment accuracy (97–98%) but does not improve generalization (86–89%). In contrast, emotion-sensitive features alone yield substantially lower performance (within-segment 89–90%; cross-segment 66–72%).

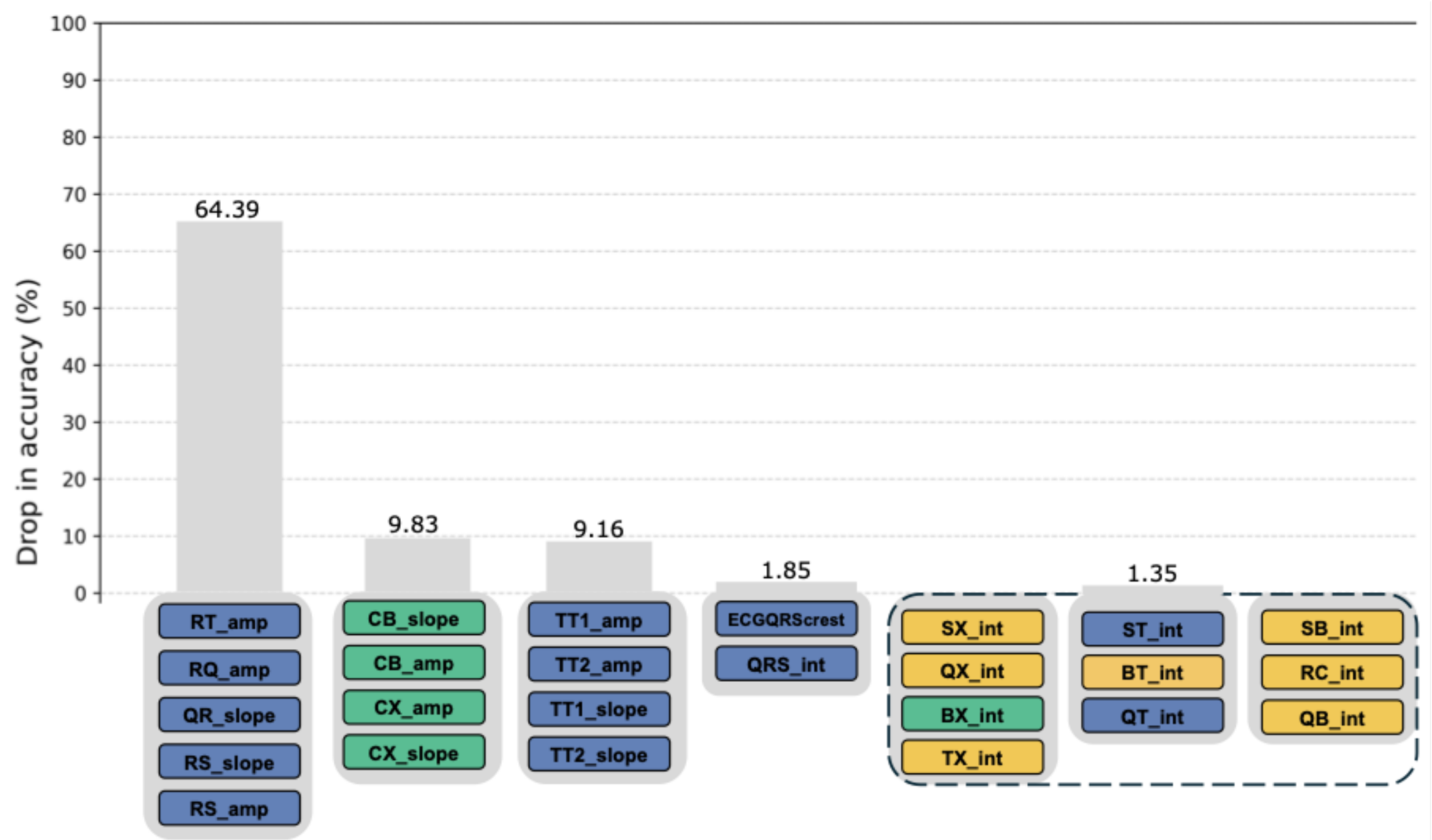

*Figure 4: Clusters of highly correlated features and the associated drop in identification accuracy (%) when each cluster is shuffled. Feature tiles are color-coded by signal of origin: dark blue = ECG only, green = ICG only, yellow = ECG\ICG. The final dashed group shows the cumulative effect of the remaining low-impact clusters.*

## 4. Discussion

The RF classifier achieves strong performance, with all evaluation metrics reaching approximately 99%, surpassing previously reported results (Antić et al., 2020; Tanasković et al., 2024). As the focus of this work is not ML optimization, but rather the analysis of feature relationships and their contributions to classification, this improvement is likely driven by several methodological differences. The number of samples per subject is increased to 18 in the present study, compared with 8 or 12 in earlier work, and the feature set is expanded from 17 to 29 features, introducing more temporal and amplitude features as well as introducing a new feature domain called slope, inspired by Patro et al. (2022) and Pandey & Keshri (2025). Interestingly, Patro et al. (2022) and Pandey & Keshri (2025) classify slope and angle as distinct feature categories, while the proposed study treats them as functionally equivalent, as angle can be directly computed from slope using trigonometric transformation, while both convey the same underlying geometric relationship between fiducial points. In addition, only baseline and anger segments are used for feature extraction, with neutral segments excluded.

### 4.1 Interpretation of correlation and clustering patterns

The correlation analyses reveal substantial interdependencies among variables. Twenty-seven feature pairs exhibit a strong Spearman correlation coefficient, indicating robust monotonic associations insensitive to outliers. Pearson correlations confirm these findings with an additional four pairs exceeding the 0.7 threshold in absolute value, suggesting that linear relationships predominate, but are complemented by monotonic non-linear patterns. Overall, these results emphasize the redundancy inherent in ECG- and ICG-derived features, reflecting the physiological interconnectedness of cardiac signals. Even with strong multicollinearity, RF performance does not deteriorate, consistent with its known robustness to correlated inputs (Breiman, 2001).

Based on the results presented in Table 2, almost all correlated features (except for RC_int) show an increase in their importance scores, and the rate of increase does not align with the strength of the correlation. This pattern is consistent with the known tendency of Gini importance to distribute importance across highly correlated variables (Gregorutti et al., 2017), since trees in the ensemble select among them inconsistently. Cluster-level shuffling further indicates that QRS-related features (RQ_amp, RT_amp, QR_slope, RS_amp, RS_slope) exhibit the highest drop, as shown in Figure 4, aligning with prior studies highlighting the central role of QRS complex in decision-making (Pandey & Keshri, 2025; Pinto & Cardoso, 2020; Silva Teodoro et al., 2017), while ST_int, QT_int, and BT_int exhibit the lowest decrease in accuracy (only 0.17%), despite their physiological importance.

### 4.2 Insights from feature importance analysis

ECG-derived features, primarily those associated with the QRS complex, appear at the top of the rankings across all importance rankings (Figures A2-A4 in the Supplementary Materials (Tanasković et al., 2025)). Repolarization-related descriptors such as RT_amp and TT2_amp also appear among the top features, indicating that T-wave morphology contributes additional individuality, though with lower stability than QRS-based metrics. This supports evidence that ECG alone provides strong discrimination (Meltzer & Luengo, 2025; Odinaka et al., 2012; Pereira et al., 2023), while the present results further indicate that the QRS complex plays a central role in the decision-making process.

ICG-derived descriptors, particularly CX_amp and CB_slope, also recur across methods, suggesting that the BCX complex of the ICG provides meaningful but less stable information than ECG features, supporting prior findings on its biometric potential, initially hypothesized by Rathore et al. (2021) and subsequently demonstrated in earlier empirical studies (Antić et al., 2020; Tanasković et al., 2024). The present results align with those of Benabdallah et al. (2026), who report 100% identification accuracy after integrating ECG, ICG, and blood pressure signals in a cohort of 30 individuals. Although the approach proposed by Benabdallah et al. (2026) is based on artificial neural networks and does not provide explicit physiological interpretability or insight into the underlying cardiovascular mechanisms, the reported performance further supports the growing recognition of ICG as a valuable biometric signal. However, the lower stability of ICG features requires further investigation, as it may be influenced by highly correlated features that exhibit opposite sensitivity to emotional variation, as shown in Table 3. Retraining the RF model using a common subset yields performance only ~2% lower than the full model with ~72% reduction in feature space, raising a practical question of balancing feature set reduction with the potential gains in storage efficiency and computational cost.

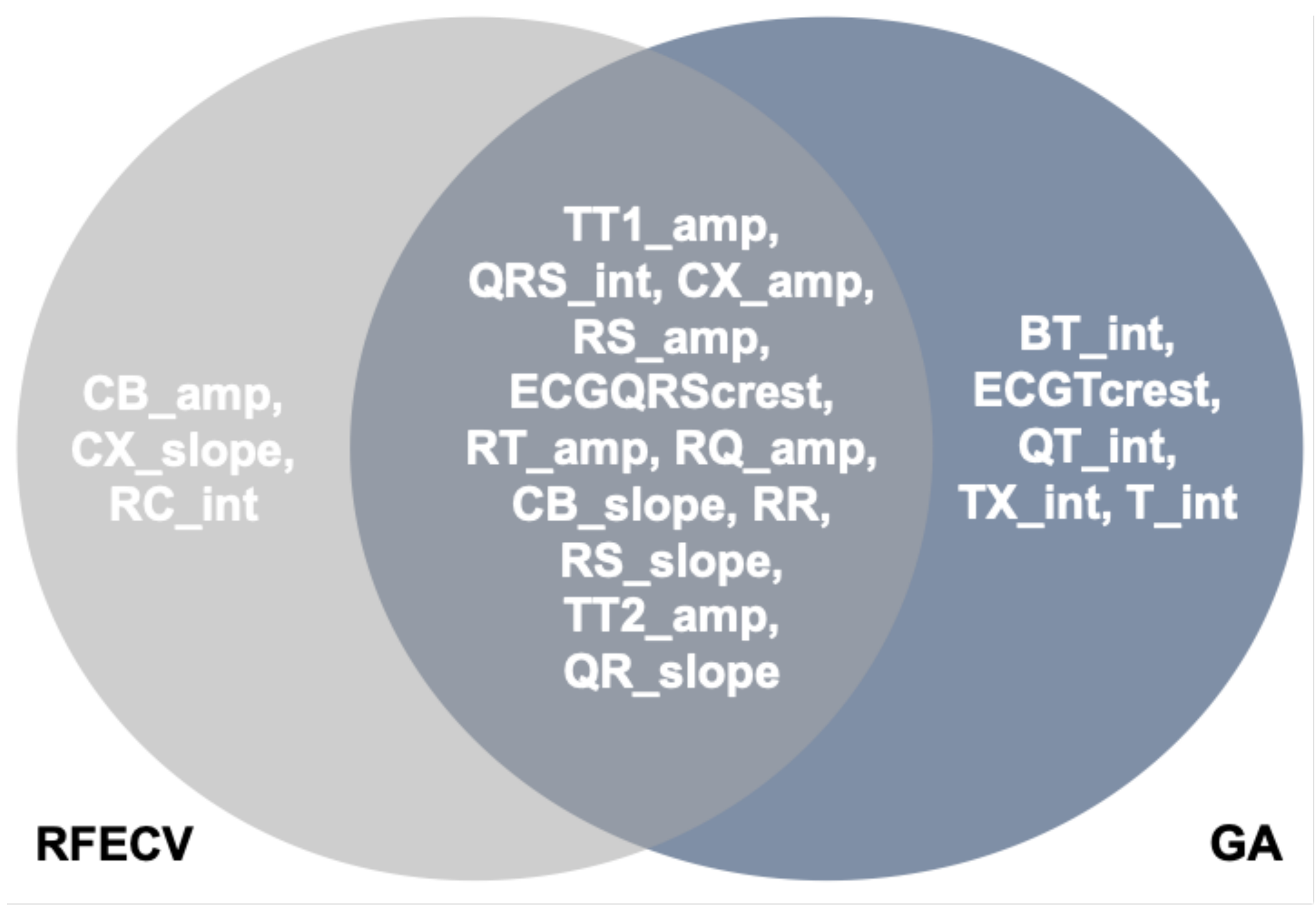

*Figure 5: Venn diagram illustrating the overlap of features selected by recursive feature elimination with cross-validation (RFECV) and genetic algorithm (GA).*

### 4.3 Evaluation of feature selection strategies

Feature selection using RFECV results in a reduced subset of 15 features while maintaining the evaluation metrics (~99%) compared to the base model. On the other hand, the GA approach selects 17 features and achieves similar performance. Two methods converge on 12 common features, shown in Figure 5, and this intersection fully overlaps with the 8 features consistently identified across all importance metrics. This further reinforces the central role of QRS-derived features in driving identification performance, as well as complementary information within the T wave. At the same time, the retention of CX_amp and CB_slope among the 12 shared features indicates that BCX features also contribute complementary information, although their absence from the SHAP-derived subset suggests that their discriminative value is less consistent across methods. These 12 common features constitute only 42% of the original feature set yet achieve performance within ~1% of the base model, underscoring that dimensionality reduction can retain nearly all discriminative power while substantially lowering computational, storage, and energy costs.

*Table 2: Impact of multicollinearity on Gini feature importance. For each permuted feature, the table reports the percentage change in the importance of its correlated features relative to their original importance. Changes across all remaining features are summarized as median and interquartile range (IQR).*

| Permuted feature | Correlated feature | Correlation coefficient | Change in importance of correlated feature (%) | Change in importance of remaining features; median (IQR) (%) |
|---|---|---|---|---|
| RQ_amp | RT_amp | 0.96 | 16.71 | 5.66 (2.46 - 7.04) |
| | QR_slope | 0.92 | 11.98 | |
| | RS_amp | 0.89 | 10.92 | |
| | RS_slope | 0.77 | 5.53 | |
| QX_int | SX_int | 0.98 | 13.47 | 1.18 (-0.89 – 4.34) |
| | BX_int | 0.94 | 7.12 | |
| | TX_int | 0.91 | 5.76 | |
| TT2_slope | TT2_amp | 0.95 | 11.49 | 2.50 (0.61 – 4.73) |
| | TT1_amp | 0.87 | 14.26 | |
| | TT1_slope | 0.79 | 9.83 | |
| CX_amp | CB_slope | 0.86 | 18.65 | 2.82 (0.32 – 5.60) |
| | CB_amp | 0.85 | 17.68 | |
| | CX_slope | 0.73 | 20.41 | |
| ST_int | QT_int | 0.92 | 3.67 | 1.79 (0.09 – 3.30) |
| | BT_int | 0.71 | 2.59 | |
| SB_int | QB_int | 0.83 | 7.36 | 2.25 (-0.90 – 3.01) |
| | RC_int | 0.77 | -0.34 | |
| QRS_int | ECGQRScrest | 0.73 | 6.79 | 3.62 (0.35 – 6.99) |

The compact feature set identified in this study is consistent with existing literature that highlights the central role of the QRS complex in ECG-based biometric identification. In particular, Pandey & Keshri (2025) applied an ensemble feature-selection approach based on three evolutionary algorithms and identified an optimal subset of 24 features. Several of their selected descriptors conceptually overlap with those obtained in our study, including QR_slope, RS_slope, the RR interval, and the area under the QRS complex, which corresponds to the ECGQRScrest feature proposed here. Among the most critical features in their ranking is the QRS angle, which, alongside RQ and ST intervals, was highlighted as one of the top three discriminative features. The significance of QRS angle lines up with earlier findings reported by Silva Teodoro et al. (2017), further emphasizing the centrality of QRS morphology. Interestingly, ST interval, identified by Pandey & Keshri (2025) as a top-ranking feature, is neither selected by any feature selection method nor ranked among the top 10 features in the present feature importance analysis. This discrepancy may reflect the influence of emotional modulation within our dataset, as the ST interval exhibits significant variation between baseline and anger conditions ($p < 0.001$), potentially reducing the stability of the ST interval as an inter-individual discriminator. Supporting this from a broader perspective, a recent study on ECG-based verification, which applied deep reinforcement learning to guide feature selection (Baek et al., 2023), arrived at remarkably similar conclusions, despite the use of advanced, non-traditional methods. Namely, QS_slope emerged as the most frequently selected feature, with RR interval, QR_slope, and RT_amp also ranking highly. While the task differs (verification rather than identification), the overlap in selected features underscores the prominence of QRS-based features, particularly those capturing slope and amplitude, across biometric paradigms.

*Table 3: Correlation and statistical significance of feature pairs. The table reports the correlation coefficient between Variable 1 and Variable 2 and indicates whether each variable differs significantly between the baseline and anger conditions.*

| Variable 1 | Variable 2 | Correlation coefficient | Variable 1 Significant | Variable 2 Significant |
|---|---|---|---|---|
| QX_int | SX_int | 0.98 | True | False |
| BX_int | SX_int | 0.95 | True | False |
| QX_int | TX_int | 0.91 | True | False |
| TT1_amp | TT2_amp | 0.90 | True | False |
| BX_int | TX_int | 0.88 | True | False |
| TT1_amp | TT2_slope | 0.87 | True | False |
| CX_amp | CB_slope | 0.87 | False | True |
| CB_amp | CX_amp | 0.85 | True | False |
| TT2_amp | TT1_slope | 0.82 | False | True |
| TT1_slope | TT2_slope | 0.79 | True | False |
| RC_int | SB_int | 0.77 | True | False |
| QRS_int | ECGQRScrest | 0.73 | True | False |

### 4.4 Interpretation of selected features from physiological and ML perspective

Inter-individual variability in the selected features plausibly arises from stable anatomical, electrophysiological, and hemodynamic differences. QRS-centric ECG features primarily capture ventricular depolarization amplitude, synchrony, and conduction velocity. These measures depend on myocardial mass and geometry, Purkinje network distribution, thoracic anatomy (heart position, chest wall thickness, lung volume), and tissue conductivity, which differ among individuals and yield unique waveform morphology (Katz, 2010; Macfarlane et al., 2010). Repolarization-related measures, which describe T-wave size and shape, reflect differences in ventricular recovery dynamics, while RR captures intrinsic sinus rate and autonomic balance, which indirectly influence morphology (Cacioppo et al., 2007). BCX descriptors index ventricular ejection (Cacioppo et al., 2007; Karpiel et al., 2022; Sherwood et al., 1990), as their amplitude and slopes depend on contractility, ejection velocity, aortic compliance, valve motion, and thoracic impedance, all of which vary across individuals. Overall, QRS-based features dominate because they arise from relatively fixed structural and conduction properties, while BCX features provide subject-specific hemodynamic information, and T-wave measures contribute individuality through repolarization gradients.

Even when the underlying physiological meaning of the QRS complex or BCX waveform is not explicitly considered, the prominence of these features in the signal, particularly the sharp and spike-like shape of the QRS complex, naturally positions them as the starting point for analysis. From a signal processing perspective, these features stand out visually and structurally, making them an intuitive anchor for segmentation and further processing. This remains the case even in modern approaches that do not rely on handcrafted features; as highlighted in the review by Meltzer & Luengo (2025), most deep learning-based studies still incorporate preprocessing and segmentation stages that use a minimal number of fiducial points, often centered around the QRS complex. Although these models do not interpret the ECG signal in physiological terms, explainable AI techniques applied by Pinto & Cardoso (2020) have shown that the focus of the network remains on the same signal landmarks, including the R peak, as well as the Q and S points. This signal-driven centrality is further illustrated in the work of Zehir et al. (2024), where identification was successfully performed using only the QRS complex, leveraging recurrent neural networks and empirical mode decomposition methods for biometric identification. These consistent findings across diverse methodologies suggest that the QRS complex continues to hold central importance in ECG-based biometric recognition, both in traditional signal analysis and in data-driven learning models.

### 4.5 Implications of emotion-sensitive cardiogram features

Previous results obtained on the same dataset (Tanasković et al., 2024) shows that emotional mismatch between enrollment and evaluation, as an example of intra-individual variation, can reduce identification accuracy by up to 13%. This result is also supported by Carvalho & Brás (2024), where a decrease in identification performance was observed when the model was trained and evaluated on samples drawn from different sessions containing distinct emotional states (neutrality, fear, and happiness). However, as the sessions were separated by at least one week, the performance degradation more broadly reflects cardiac variability introduced by emotional, physiological, and instrumental factors, such as changes in electrode placement or participant state (Carvalho & Brás, 2024). On the other hand, earlier findings by Israel et al. (2005) reported somewhat conflicting results – that anxiety did not significantly affect ECG-based authentication. Similarly, Zhou et al. (2021) found minimal impact of psychological stress on ECG biometric performance. The methodological differences and variability in feature sets across prior studies motivate approaches that explicitly clarify the impact of emotional state at the feature level, as pursued in this study, to better understand cardiographic components and their underlying psycho-physiological associations, rather than treating biometric identification as a purely data-driven task.

In the present analysis, 14 features appear as emotionally sensitive, without a clear pattern regarding domain or signal of origin. Notably, the QRS-derived features (RQ_amp, RS_amp, QR_slope, RS_slope, ECGQRScrest), which are most influential for decision-making, show no significant differences between emotional states, indicating that the core identity-relevant descriptors remain stable.

Training on only emotion-insensitive features yields a further reduction in cross-emotion performance, indicating that removing emotion-sensitive features does not improve robustness. In contrast, models trained exclusively on emotion-sensitive features show larger degradation in performance, demonstrating that emotion-driven variability provides little subject-discriminative value. This suggests that emotionally sensitive features carry limited information on individuality, which is expected given that identity-specific descriptors must remain stable across varying emotional states for a biometric system to function reliably. Although Bazett's correction is applied, several intervals still differ significantly, suggesting that additional normalization strategies may be useful.

Interestingly, when using all or only emotion-insensitive features, cross-segment generalization is consistently better from anger to baseline than the other way around, which is rather unexpected given that baseline is typically considered a segment with no influence.

*Table 4: Random forest performance across baseline and anger segments using all features, non-significant features showing no statistically significant baseline–anger difference, and significant features showing a statistically significant baseline–anger difference. Cross-validation (CV) refers to training and evaluation within the same segment, while generalization (Gen) refers to training on one segment and testing on the other. Acc - accuracy*

| Training segment → Evaluation segment | Setting | All features | | Non-significant | | Significant | |
|---|---|---|---|---|---|---|---|
| | | Acc (%) | F1 (%) | Acc (%) | F1 (%) | Acc (%) | F1 (%) |
| Baseline → Baseline | CV | 98.62 | 98.53 | 97.74 | 97.60 | 89.99 | 89.12 |
| Baseline →Anger | Gen | 89.44 | 88.16 | 87.73 | 86.31 | 72.11 | 69.43 |
| Anger → Anger | CV | 97.63 | 97.36 | 97.96 | 97.78 | 90.54 | 89.83 |
| Anger → Baseline | Gen | 92.13 | 91.22 | 89.00 | 87.54 | 69.58 | 66.82 |

Beyond the stable QRS descriptors, T-wave and ICG BCX features are split between significant and non-significant sets, as shown in Table 3, which could cause their lower stability and smaller overall contribution to identification. Overall, the results indicate that not all features provide equivalent identity-specific information, and optimal feature choice depends on the intended goal: enhancing biometric performance or mitigating emotional effects.

The present findings suggest that cardiographic signals should be interpreted as carriers of multiple dimensions of information about individuals. First, they contain relatively stable anatomical, electrophysiological, and hemodynamic properties that can support biometric identification. Second, ECG-derived parameters, particularly those related to autonomic regulation, may also reflect relatively stable psychological characteristics, allowing ML-based models to infer personality traits from ECG features (Boljanić et al., 2022). This is consistent with broader personality-assessment research showing that ML can extract psychologically meaningful information from high-dimensional human data, while also requiring careful interpretation of what the model has actually learned (Stachl et al., 2020). In contrast to these more stable dispositional or identity-related signals, emotional responses reflect transient and context-dependent physiological modulation. Therefore, cardiogram-based identification requires distinguishing between features that act as stable carriers of individuality and features that mainly reflect momentary affective state. The present results support this distinction: QRS-derived descriptors appear to behave as robust identity carriers, whereas several temporal, T-wave, and ICG-related features show greater sensitivity to anger and contribute less consistently to identification. This distinction between stable identity information, complementary cardiographic cues, and temporal features sensitive to emotional change is summarized in Figure 6.

This distinction is important for the development of physiological authentication systems intended for realistic interactive contexts. Users may attempt authentication under different emotional, cognitive, and situational conditions, and reliable systems should therefore rely on features that preserve identity-specific information despite such variability (Belk et al., 2017; Pereira et al., 2023). The use of multiple interpretability methods supports this goal by clarifying whether model decisions are based on physiologically meaningful and task-appropriate signal components rather than redundant statistical patterns. This aligns with the broader view that interpretable ML should be evaluated in relation to the purpose of the decision and the consequences of relying on the model output (Doshi-Velez & Kim, 2017). Therefore, the compact 12-feature subset identified here is relevant not only because it preserves near-complete identification performance with reduced complexity, but also because it provides a more interpretable set of cardiographic descriptors for future studies of stable person-related characteristics and state-dependent physiological variability.

Stable identity information

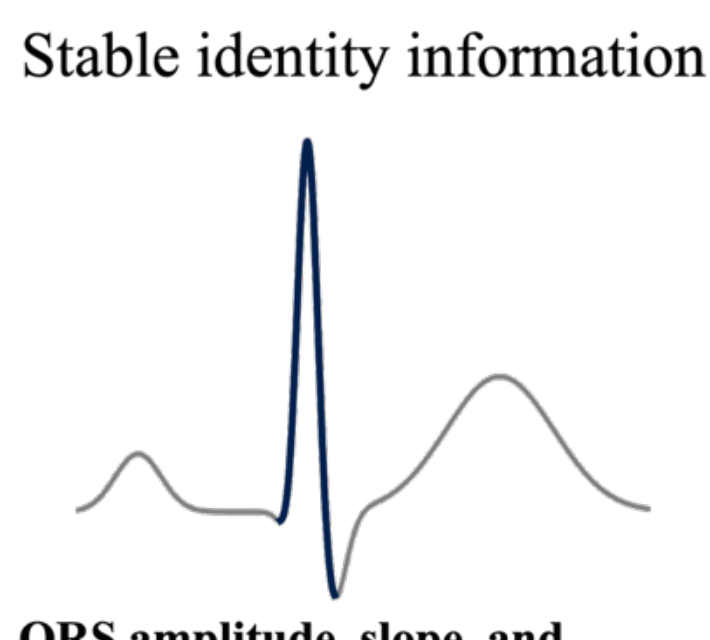

**QRS amplitude, slope, and morphology**
Stable biometric information
Stable across emotional conditions

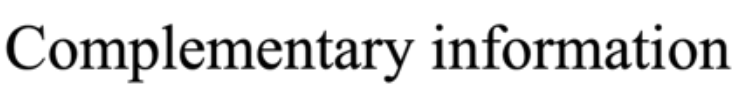

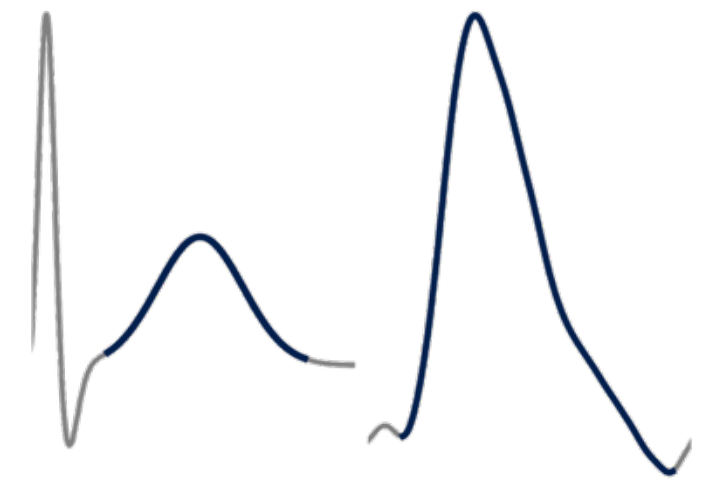

**T wave and BCX descriptors**
Contribute with complementary identity information
Show mixed stability across methods and emotions

Transient features

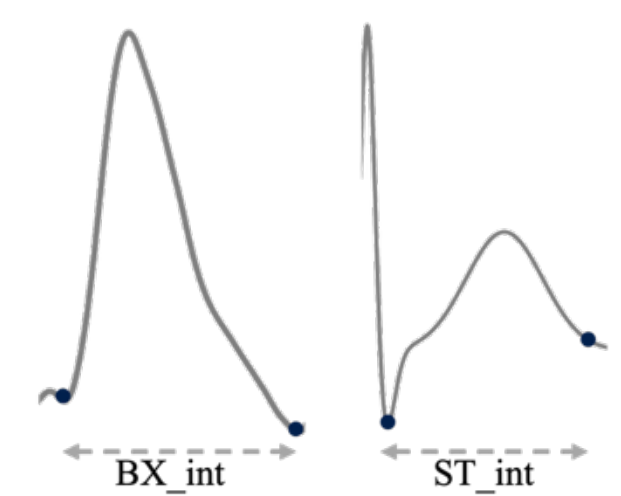

**Intervals and durations**
Change between baseline and anger
Reflect transient physiological variation

*Figure 6: Conceptual summary of cardiographic information. QRS amplitude, slope, and morphology represent stable identity carriers, whereas T wave and BCX descriptors provide complementary but less consistent information. Representative temporal intervals illustrate features that are more sensitive to emotional change.*

### 4.6 Limitations of the study

Here, we recognize the limitations of the proposed study:

1. Future work should include additional fiducial descriptors, such as P-wave features and more detailed ICG delineation, not to further increase already high accuracy but to provide a more comprehensive exploratory mapping of individual-specific information across the cardiac cycle. However, these descriptors may introduce additional uncertainty because P-wave detection is challenging and ICG fiducial-point detection is sensitive to noise and waveform variability (Maršánová et al., 2019; Trybek et al., 2023).
2. The present analysis is limited to baseline and anger. Future studies should examine a broader range of emotional states, as different emotions may modulate the relevant cardiographic features in distinct ways and may therefore affect the stability of identity-related information differently (Carvalho & Brás, 2024; Israel et al., 2005; Zhou et al., 2021).
3. The 2-minute segments are too short for valid spectral HRV estimation. Guidelines recommend ≥ 1–2 minutes, preferably (5 – 20) minutes, of stationary data for LF/HF analysis (Camm et al., 1996; Damoun et al., 2024). Accordingly, frequency-domain HRV metrics are not extracted.
4. The dataset shows an unbalanced sex distribution reflecting the psychology-student population (Odic & Wojcik, 2020; Olos & Hoff, 2006). This imbalance is not corrected, as sex is not used for stratification, and potential sex-related differences in identification accuracy are not examined.

5. The dataset consists of healthy young adults, which may limit generalizability to broader user groups differing in age, cultural background, technology familiarity, and psychological or clinical characteristics (Belk et al., 2017; Pereira et al., 2023). Future studies should test whether the identified identity carriers remain stable across more heterogeneous populations and interaction contexts.
6. Future work should explore alternative normalization strategies (*e.g.*, Israel et al., (2005) to evaluate their effect on model performance.
7. The neutral condition is not used in the current analysis, although it may serve future investigations on cognitive-workload effects on cardiographic signals, as suggested by Pale et al. (2021).

### 4.7 Contributions of the study

Taking into consideration all of the data provided in this study, we now revisit the initial research questions and present a summary of the findings in relation to each of them:

1. The most reliable information comes from features derived from the QRS complex. Across three feature importance methods, QRS features (ECGQRScrest, RQ_amp, RS_amp, QR_slope, RS_slope, QRS_int) rank consistently high. These patterns align with physiology: QRS amplitude and slope reflect stable structural and conduction properties, whereas BCX features capture inter-individual differences in ejection dynamics, aortic compliance, and valve timing.
2. The RF model inherently adjusts importance scores for correlated features, confirming their reduced individual contributions. The study shows that almost all highly correlated features (except RC_int) shows an increase in importance scores upon shuffling, but the rate of increase does not always align with their correlation coefficients.
3. QRS-based features (except QRS_int) remain stable during anger, consistent with their negligible effect sizes. Predictive analysis mirrors this: using only emotion-insensitive features does not improve cross-emotion generalization (accuracy drops ~11%). Using only statistically significant features further reduces cross-emotion accuracy by up to ~23%, indicating that emotion-driven variance does not support reliable identification. Together, these findings show that QRS-derived features retain high identity specificity across emotional states, whereas emotion-sensitive features degrade generalization.

## 5. Conclusions

This study proposes a comprehensive exploratory framework for interpreting the decision-making processes inherent in cardiogram-based biometric identification. By shifting the focus from predictive performance to an explainable analytical pipeline, we address a gap in the deployment of physiological biometrics, which is the transition from "black-box" models to transparent, biologically grounded systems or at least as transparent and as grounded as possible. Our approach, which integrates feature importance, feature selection, cross-correlation analysis, and evaluation of emotional effects, demonstrates that identity-related information is not uniformly distributed across the cardiac cycle but is heavily concentrated within the QRS complex of the ECG.

From a physiological perspective, the dominance of amplitude- and slope-based descriptors in the QRS complex suggests that the most discriminative features are those tied to fixed anatomical invariants, such as ventricular mass, conduction pathways, and thoracic geometry. Additionally, features derived from the BCX complex of the ICG provide additional discriminative information, supporting a multimodal approach, but exhibit lower stability across analyses, indicating a secondary role in identification, reflecting variations in ventricular ejection dynamics and characteristics of the vascular system.

The analysis further shows that cardiographic feature spaces are highly redundant, with strong correlations among features forming seven distinct clusters. As a consequence, RF distributes importance scores across features within clusters by using them interchangeably and underestimating the actual importance of individual features. By applying an ensemble of GA and RFECV, the results show that this high-dimensional manifold can be projected onto a compact subset of 12 features (out of 29), while maintaining a performance margin within 1% of the full feature set (99%). This confirms that efficient, real-time biometric models can be achieved through strategic dimensionality reduction without sacrificing the underlying discriminative signal.

Finally, the comparison between baseline and anger shows that the principal identity information carried by QRS amplitude, slope, and morphology remains stable across emotional conditions, while QRS duration was sensitive to emotional change. In contrast, features related to the T wave and the BCX complex show mixed responses, with some remaining stable and others changing significantly. This variability may explain why these components provide complementary identity information but show lower consistency than the main QRS descriptors. The poor generalization achieved using only emotion-sensitive features further indicates that transient physiological variation provides limited reliable information about identity. While the generalizability to other high-arousal emotions remains an open question, this research provides a blueprint for linking ML decisions to physiologically meaningful components. By prioritizing transparency and explainability, the proposed framework supports a clearer understanding of which cardiographic components carry stable individual-specific information and which are more sensitive to transient affective changes. In this way, the study contributes not only to cardiogram-based identification, but also to future psychological and psychophysiological applications in which cardiographic signals can be used to study stable person-related characteristics and state-dependent variability.

**CRediT author contributions statement**

**Ilija Tanasković:** Methodology, Software, Formal analysis, Visualization, Writing – Original Draft **Ljiljana B. Lazarević:** Conceptualization, Methodology, Data Curation, Writing - Review & Editing **Goran Knežević:** Conceptualization, Resources, Investigation, Writing - Review & Editing **Nikola Milosavljević:** Data Curation, Methodology, Investigation, Visualization, Writing - Review & Editing **Olga Dubljević:** Data Curation, Methodology, Investigation, Writing - Review & Editing **Bojana Bjegojević:** Data Curation, Methodology, Investigation, Writing - Review & Editing **Nadica Miljković:** Conceptualization, Methodology, Validation, Formal analysis, Writing - Review & Editing.

**Funding**

Nadica Miljković acknowledges the support from the Ministry of Science, Technological Development and Innovation of the Republic of Serbia [Grant No. 451-03-34/2026-03/200103]; Ljiljana B. Lazarević and Goran Knežević acknowledge the support from the Ministry of Science, Technological Development, and Innovation of the Republic of Serbia [Grant No.  451-03-66/2025-03/200163]; Nikola Milosavljević acknowledges the support from the Ministry of Science, Technological Development and Innovation of the Republic of Serbia [Grant No.

451-03-137/2025-03/200096]. The funder was not involved in the manuscript preparation and the decision to submit the manuscript. Also, the funder was not related to the study design, data collection, data analysis, or results interpretation.

**Declaration of competing interests**

The authors declare that they have no known competing financial interests or personal relationships that could have appeared to influence the work reported in this paper. The Funder did not participate in any aspect of the study design, collection, analysis, and interpretation of data; in the manuscript preparation; or in the decision to submit the manuscript.

**Generative AI disclosure statement**

During the preparation of this work, the Authors used GPT-5 (ChatGPT) to improve readability and language. After using this tool/service, the Authors reviewed and edited the content as needed and take full responsibility for the content of the publication.

No AI-generated or AI-altered images were used. All figures were created by the Authors.

**Ethics statement**

All participants signed Informed Consent forms in accordance with the Helsinki Declaration, and the Institutional Review Board from the Department of Psychology at the University of Belgrade approved the study (No. 2018-19) on December 5, 2018.